\documentclass[5p,twocolumn]{elsarticle}
\usepackage{graphicx,latexsym}
\usepackage{dcolumn}
\usepackage{amssymb,amsmath,bm}
\usepackage{subfigure}
\usepackage{braket}
\usepackage{multicol}
\usepackage{siunitx}
\usepackage{array}
\usepackage{float}
\usepackage[nopar]{lipsum}
\usepackage{caption}
\usepackage{multirow}
\usepackage{bigstrut}
\usepackage[super]{nth}

\newcommand{\angstrom}{\text{\normalfont\AA}}
\usepackage{setspace}
\usepackage{hyperref}
\hypersetup{
    pdfnewwindow=true,       % links in new window
    colorlinks=true,         % false: boxed links; true: colored links
    linkcolor=blue,          % color of internal links
    citecolor=blue,          % color of links to bibliography
    filecolor=magenta,       % color of file links
    urlcolor=black           % color of external links
}

\usepackage[normalem]{ulem}

\def\sec#1{Sec.\ \ref{#1}}

\def\fig#1{Fig.\ \ref{#1}}
\def\tab#1{Tab.\ \ref{#1}}

\journal{}

\begin{document}

\begin{frontmatter}

%-----------------------------------------------------------------

\title{Interlayer interaction controlling the properties of AB- and AA-stacked bilayer graphene-like BC$_{14}$N and Si$_{2}$C$_{14}$}

%	\title{Outstanding performance of electronic, mechanical, optical and thermal properties of
%	bilayer graphene-like BC$_{14}$N and Si$_{2}$C$_{14}$ semiconductor materials}

\author[a1,a2]{Nzar Rauf Abdullah}
\ead{nzar.r.abdullah@gmail.com}
\address[a1]{Division of Computational Nanoscience, Physics Department, College of Science, 
             University of Sulaimani, Sulaimani 46001, Kurdistan Region, Iraq}
\address[a2]{Computer Engineering Department, College of Engineering, Komar University of Science and Technology, Sulaimani 46001, Kurdistan Region, Iraq}

\author[a1]{Hunar Omar Rashid}

\author[a5]{Andrei Manolescu}
\address[a5]{Reykjavik University, School of Science and Engineering,
	Menntavegur 1, IS-101 Reykjavik, Iceland}

\author[a6]{Vidar Gudmundsson}   
\address[a6]{Science Institute, University of Iceland,
	Dunhaga 3, IS-107 Reykjavik, Iceland}

%
%----------------------------------------------------------------

\begin{abstract}
\onehalfspacing
We model bilayer graphene-like materials with Si$_2$C$_{14}$ and  BC$_{14}$N stoichiometry, where the interlayer interactions play important roles shaping the physical properties of the systems. 
We find the interlayer interaction in Si$_2$C$_{14}$ to be repulsive due to the interaction of Si-Si atoms, and in BC$_{14}$N it is attractive due to B and N atoms for both the AA- and the AB-stacking.
The repulsive interlayer interaction opens up a bandgap in Si$_2$C$_{14}$ while 
the attractive interlayer interaction in BC$_{14}$N induces a small indirect bandgap or overlaping of the valence conduction bands. Furthermore, the repulsive interaction decreases the Young modulus while the attractive interaction does not influence the Young modulus much. The stress-strain curves of both the AA- and the AB-stackings are suppressed compared to pure graphine bilayers.
The optical response of Si$_2$C$_{14}$ is very sensitive to an applied electric field and an enrichment in the optical spectra is found at low energy. The enrichment is attributed to the bandgap opening 
and increased energy spacing between the $\pi{\text -}\pi^*$ bands.
In BC$_{14}$N, the optical spectra are reduced due to the indirect bandgap or the overlapping 
of the $\pi{\text -}\pi^* $ bands. Last, a high Seebeck coefficient is observed due to the presence 
of a direct bandgap in Si$_2$C$_{14}$, while it is not much enhanced in BC$_{14}$N. 
\end{abstract}


\end{frontmatter}

\section{Introduction} 
\onehalfspacing
The ability to control physical properties of a material by tuning its bandgap is at the heart of modern electronic devices \cite{Alattas2018}. Studies have found that an energy gap can 
be generated in two-dimensional materials, such as a monolayer \cite{Elias2019} or a bilayer grahene (BLG) \cite{doi:10.1021/nn202463g}. 
BLG consisting of two stacked monolayers of graphene is a material in which physical properties 
can be controlled by tuning the bandgap \cite{Novoselov666}.
It is thus possible to improve the electrical, the thermal and the optical conductivities and 
modulate the mechanical properties \cite{McCann_2013}. 
However, several of the characteristics of BLG are similar to those of a monolayer, 
but a BLG has interesting underlying physics that
holds a potential for electronics applications such as sensors \cite{SEEKAEW2017357}, exhibiting high sensitivity \cite{QIN2017760}, stable specificity, and fast response \cite{PhysRevB.100.075421}.

The electronic properties of a BLG have been studied using density functional theory (DFT), in which 
the bandgap at the $K$ point in the Brillouin zone depends linearly on the average
applied electric field \cite{NEMNES20199, PhysRevB.79.165431}. One of the main approaches to alter the electrostatic potential of a BLG is substitutional doping with foreign atoms \cite{C0JM02922J} such as Boron (B), Nitrogen (N) \cite{NEMNES2018175}, and Silicon (Si) atoms. For instant, a B- and N-doped BLG results in a p-type or an 
n-type semiconducting behavior with a shifting of the Fermi energy, respectively. A B and N codoped 
BLG exhibits a semiconding material with a small bandgap, where the Fermi energy is located in 
the bandgap \cite{doi:10.1002/adma.200901285}. 
A substantial bandgap due to Boron and Nitrogen atoms codoped in BLG can be created in which the size of the band gap is effectively tuned in the presence of B-N pairs \cite{Alattas2018}.
The created band gap of a boron nitride (BN) BLG by varying an external electric field can be used for investigating photocatalysis \cite{doi:10.1063/1.4950993}. 
These electronic properties of B or N doped BLG give almost similar result for
both the AA- and AB-stackings patterns with respect to the dopant atoms 
\cite{doi:10.1021/acs.nanolett.9b00986}.

The doping of graphene with Si, N and B atoms leads to modified bonds in the BLG which in turn effects its mechanical properties.  
The influence of the Si, N and B doping on the mechanical properties of graphene has been examined
and it was shown that it leads to an almost linear decrease of the Young’s modulus. Such doping
effects are found to be most significant for silicon, less pronounced for boron, and small or negligible for nitrogen \cite{HAN2015618}. In addition, molecular dynamics simulations have been done for twisted bilayer of graphene displaying that they possess outstanding mechanical properties. 
In the linear elastic region, the mechanical strain rate and the presence of cracks have 
negligible effects, while in the nonlinear mechanical region fracture toughness is seen \cite{liu2018molecular}.

The optical characteristics of BLG are interesting for optoelectronics devices and graphene-based photodetectors \cite{Echtermeyer2011}. Inter- and intraband transitions in graphene have 
been studied using infrared spectroscopy finding a strong intraband absorption in the terahertz
frequency range \cite{5951298, Abdullah_2019, Abdullah2019}. The foreign atoms can be used to control the optical properties 
as a function of the doping as has been demonstrated in an electrostatically gated 
BLG \cite{PhysRevB.79.115441}.
The roles of the doping atoms on the $\pi-\pi$ interactions was studied and a systematic red shift 
and a broadening of the lowest excitations in the optical absorption was demonstrated \cite{doi:10.1021/jp504222m}. Furthermore, the Si doping opens the band gap of graphene and enhances its optical conductivity \cite{Houmad2015, en12061082}. 

The pure monolayer and bilayer graphene structures are generally not good candidates for thermoelectric based devices because of the vanishing bandgap \cite{Dollfus_2015, ABDULLAH2018223}.  In the pure material the 
Seebeck coefficient, $S$, and the thermoelectric figure of merit, $ZT$, are thus very limited. 
In order to enhance $S$ and $ZT$, one may thus introduce B and N doping atoms in the graphene structures. 
It has been shown that graphene/BN heterostructures provide the possibility to
tune and decrease strongly the phonon thermal conductivity, which is a very favourable feature to enhance the thermoelectric properties such as $S$ and $ZT$ \cite{PhysRevB.86.115410, PhysRevB.84.205444, ABDULLAH2020103282}.

In this work, we consider Si, B, and N atoms doped AA- and AB-stacked BLG represented by BC$_{14}$N and Si$_{2}$C$_{14}$ structures. The electronic, mechanical, optical and thermal characteristics are investigated using density functional theory. A comparison between the AA- and the AB-stacked BLG 
will be shown with detailed mechanisms of improving these two BLG structures by Si and BN impurity
atoms.

In \sec{Sec:Model} the structure of BLG is briefly over-viewed. In \sec{Sec:Results} the main achieved results are analyzed. In \sec{Sec:Conclusion} the conclusion of the results is presented.

%%%%%%%%%%%VG%%%%%%%%%%%%%
\section{Computational details}~\label{Sec:Model}
\onehalfspacing
The present results have been worked out by using the Quantum Espresso (QE) package \cite{Giannozzi_2009}, 
which encompasses tools for first-principles calculations and materials modeling for electronic structure
simulations based on DFT. 
A full structure relaxation is obtained by including van der Waals interactions in 
the exchange (XC) functional of the DFT model, where the $k$-point grid is $12\times12\times1$ \cite{PhysRevB.82.081101}.  
The Perdew-Burke-Ernzerhof functional (PBE) within the framework of the generalized gradient approximation is employed for the calculations of geometry optimizations and the electronic properties \cite{PhysRevB.23.5048}. 
For the Brillouin zone sampling and the calculations of the density of state (DOS), a $12\times12\times1$ and a $77\times77\times1$ grids are used, respectively.
The energy cutoff for the plane wave expansion is set to be $1088.45$ eV for all the calculations. 
Pure BLG and doped BLG graphene structures are iteratively optimized and the calculations are considered 
converged when the force on each atom is less than $10^{-6}$ eV/$\angstrom$.

The crystalline and molecular structure visualization program (XCrySDen) 
is employed to visualize all the structures \cite{KOKALJ1999176}. 
In addition, the Boltzmann transport properties software package (BoltzTraP) is used to investigate the thermal properies of the systems \cite{madsen2006boltztrap-2}. The BoltzTraP code uses a mesh of band energies and has
an interface to the QE package \cite{ABDULLAH2020126578}. The optical characteristics of the systems are obtained 
by the QE code.

\hfill

\section{Results}~\label{Sec:Results}
We approach the AA- and AB-stacked BLG with a $2\times2$ supercell. Two types of doping are considered: 
Si doped BLG and BN-codoped BLG. These two types of atom doping are expected to form BLG-like 
semiconductor materials.   

\subsection{AA- and AB-stacked structures}

\begin{figure*}[htb]
	\centering
	\includegraphics[width=0.7\textwidth]{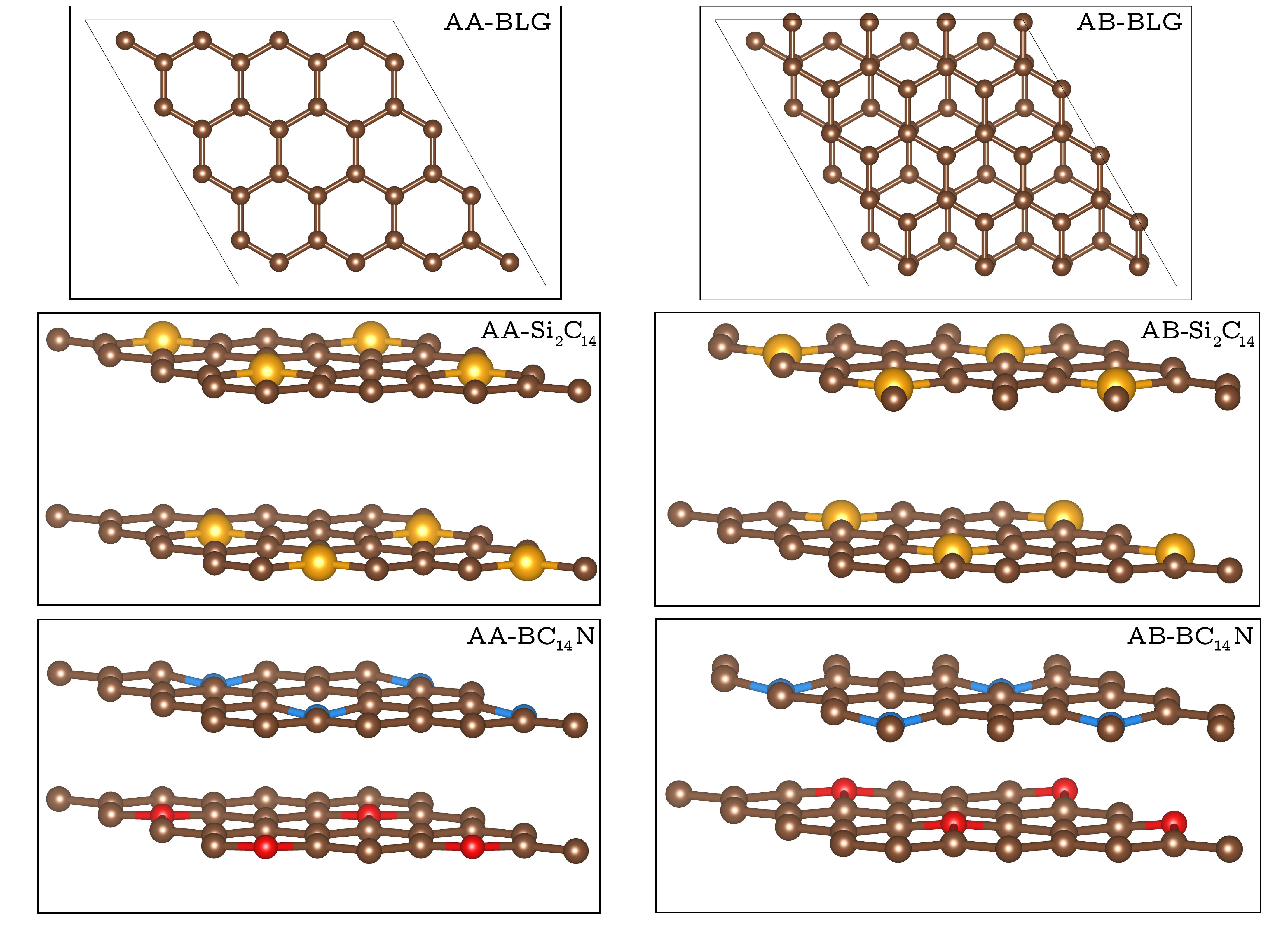}
	\caption{AA- and AB-stacked pristine BLG, Si$_2$C$_{14}$, and BC$_{14}$N. The C, Si, B and N atoms are brown, golden, blue and red colors, respectively.}
	\label{fig01}
\end{figure*}

We consider an AA- and an AB-stacked BLG in our study as are shown in \fig{fig01}. The AA-stacked 
BLG is composed of two layers that are exactly aligned, while in the AB-stacked BLG (Bernal stacking), 
the carbon atoms belonging to different sublattices, A and B, form the AB stacking pattern between 
the layers (atoms belonging to the A sublattice in one layer are stacked directly above the atoms of the B sublattice from the other layer). 
\lipsum[0]

\lipsum[0]
\begin{table*}[h]
	\centering
	\begin{center}
		\caption{\label{table_1} The lattice constant, a, the inter-layer distance, $l$, the B-N distance (d$_{\rm BN}$), the S-Si distance (d$_{\rm Si-Si}$), the C-C, B-N, C-B, and C-N bond lengths for all pure and doped structures. The unit of all parameters is $\angstrom$.}
		\begin{tabular}{|l|l|l|l|l|l|l|l|l|l|}\hline
			Str.	        & a     &  $l$   & d$_{\rm Si-Si}$& d$_{\rm BN}$& C-C& C-B& C-N & C-Si    \\ \hline
			\multicolumn{1}{| c }{AA-stacking}                                                        \\ \hline
			BLG	            & 2.46  &  3.6   &  -        &   -     &1.42     &  -     & -    &  -     \\
			Si$_2$C$_{14}$  & 2.73  &  4.5   &  5.25     &   -     &1.47     &  -     & -    & 1.69   \\		
			BC$_{14}$N      & 2.48  &  3.05  &  -        &   3.71  &1.419    &  1.48  & 1.43 &  -     \\ \hline
			\multicolumn{1}{| c }{AB-stacking}                                                         \\ \hline
			BLG	            & 2.46  &  3.4   &  -        &   -     &1.42     &  -     &  -   &        \\
			Si$_2$C$_{14}$  & 2.84  &  4.19  &  4.46     &   -     &1.44     &  -     &  -   & 1.69   \\		
			BC$_{14}$N	    & 2.51  &  2.9   &   -       &   2.94  &1.425    &  1.48  & 1.44 & -      \\ \hline
		\end{tabular}
	\end{center}
\end{table*}

Based on interaction effects between atoms or the interlayer interaction,
we consider two types of doping atoms arrangements in the BLG with Si dopant or BN-codopant atoms.
The Si-doped BLG identified as Si$_2$C$_{14}$ for both the AA- and the AB-stackings shown 
in \fig{fig01} (Middle panel). 
In addition, a BN-codped BLG that is labeled as BSi$_{14}$N for the AA- and the AB-stackings. In a $2\times2\times1$ super-cell of Si$_2$C$_{14}$, one Si (golden) atom in each layer is doped while in BSi$_{14}$N we assume one B (blue) atom in the top layer and one N (red) atom in the bottom layer. The Si atom in top(bottom) layer is doped in the para(meta) position, while the B(N) atom is located at the ortho(para) position \cite{ABDULLAH2020126350,ABDULLAH2020126807,ABDULLAH2020103282}.  

The interlayer distance of both the AA- and the AB-stacked BLG are found to be $3.6$ and $3.4$~$\angstrom$,
respectively, which are a good agreement with experimental \cite{doi:10.1063/1.2975333} and theoretical 
\cite{Alattas2018} results.
The lattice constant, a, the interlayer distance, $l$, the distance between the B-N (d$_{\rm BN}$), 
the distance between the Si-Si atoms (d$_{\rm Si-Si}$), the C-C, C-B, C-N and the C-Si bond lengths are all
presented in \tab{table_1}.
It is known that the repulsive force between the two layers in the case of the AA-stacking is much stronger than that in the AB-stacking. This is attributed to the fact that half of the atoms of the first layer are located in the center of hexagon of the second layer, and the other half of the atoms of the first layer lies directly above 
atoms of the second layer in the AB-stacking.  Therefore, a higher binding energy is required for the AA-stacking compared to the AB-stacking. Based on that, the AB-stacking is a more stable arrangement compared to the AA-stacking.
The repulsive interaction between the layers of the AA-stacking leads to a larger interlayer distance.

Another observation of the data in \tab{table_1} is that the average lattice constant of both Si$_2$C$_{14}$, and BC$_{14}$N is larger than that of BLG in both cases of an AA- and an AB-stacking indicating a super-cell expansion 
due to the dopant atoms. This is attributed to the larger atomic radii of the B and Si atoms compared to the C atom. 
In addition, the super-cell expansion for the doped AB-stacked structure is slightly larger than that of the 
AA-stacked. This may be referred to the interlayer interaction which is stronger for the AB-stacked doped systems
as the interlayer distance of AB-stacked is smaller.

Furthermore, the inter-layer distance in Si$_2$C$_{14}$ is large, but in BC$_{14}$N it is small compared to BLG for 
both the AA- and AB-stackings. This reveals a repulsive interaction between layers in 
Si$_2$C$_{14}$ \cite{DENIS2010251}, but an attractive interaction between the layers in BC$_{14}$N. 
These repulsive and attractive interactions arise due to presence of the dopant atoms.

Several methods have been used to analyse the components or the sources of the interlayer interactions in BLG. 
First, the interaction that emerges due to the sp$^3$ bonding between the two layers of BLG. 
This type of interaction is studied by incrementally moving two atoms with the same planar coordinates 
(one in the upper layer and the other one in the lower layer) to the sp$^3$ bond distance which
is about $1.54$~$\angstrom$ \cite{doi:10.1063/1.4740259}. 
Second, non-bonding potentials are used to describe the van der Waals interactions between the layers of BLG \cite{PhysRevB.62.13104, PhysRevB.80.245424, PhysRevB.85.245430}. The Van der Waals interaction is a dipole-dipole interaction that is effective if the distance between dipoles is around 4-5~$\angstrom$. Third, the non-bonding interaction energy between the layers of a BLG \cite{doi:10.1021/jp2095032}. In this case the interaction energy 
determines the type of the interlayer interaction that could be either a repulsive or an attractive interaction.

In our work, the non-bonding interlayer interaction energy between the layers of the BLG is considered, where 
the dopant atoms play an essential role \cite{Dappe_2012}.
The interaction energy between the dopant atoms in graphene can be determined or obtained from the total energy 
of the system using DFT calculation. The interaction energy between two dopant atoms in a structure can be defined as 
\begin{equation}
\Delta E = E_2 - E_0 + 2 \times E_1,
\end{equation}
where E$_0$, E$_1$ and E$_2$ are the total energies of the systems with
zero, one and two substitutional dopant atoms, respectively. 
The interaction energy between the Si atoms in the different layers of the AA- and that AB-stacked Si$_2$C$_{14}$ is found to be $2.11$~eV and $1.34$~eV, respectively, revealing a stronger repulsion interaction between the two substitutional Si atoms in the AA-stacking. The interaction is repulsive as the interaction energy has a positive value in both cases. 

Even though the Si atoms in the AA-stacking do not lie directly above each other, the repulsive inteaction is still 
stronger between the layers just like in the pure AA-stacked BLG.
We can say that the AB-stacked Si$_2$C$_{14}$ is more stable than AA-stacked because the repulsive interaction in 
the AB-stacked is smaller leading to a smaller binding energy.
This interactions between the Si atoms in a monolayer~\cite{Wei2014} and a bilayer~\cite{DENIS2010251} 
graphene have been studied an a repusive interaction between the Si atoms has been confirmed. 
 
In addition, an attractive interaction between the B and the N atoms is observed in both the AA- and the AB-stacked  BC$_{14}$N as the interaction energy has a negative value, $-3.2$, and $-2.5$~eV, respectively. 
The attractive interaction between the B and the N atoms leads to a decrease the the interlayer distance which is $3.05$~$\angstrom$ for AA-stacking and $2.9$~$\angstrom$ for AB-stacking. Furthermore, the the B atom in the top 
layer and the N atom in the bottom layer are slightly moved toward each other indicating an attractive interaction between theses two atoms. We conclude that the AB-stacking is more stable because of higher attractive interactions.
The attractive interaction between the B and the N atoms in graphene has been studied in two dimentional systems 
such as in monolayer graphene and in silicene nanosheet~\cite{doi:10.1063/1.4742063, abdullah2020properties}.

\subsection{Electronic Band structure}

\begin{figure}[htb]
	\centering
	\includegraphics[width=0.45\textwidth]{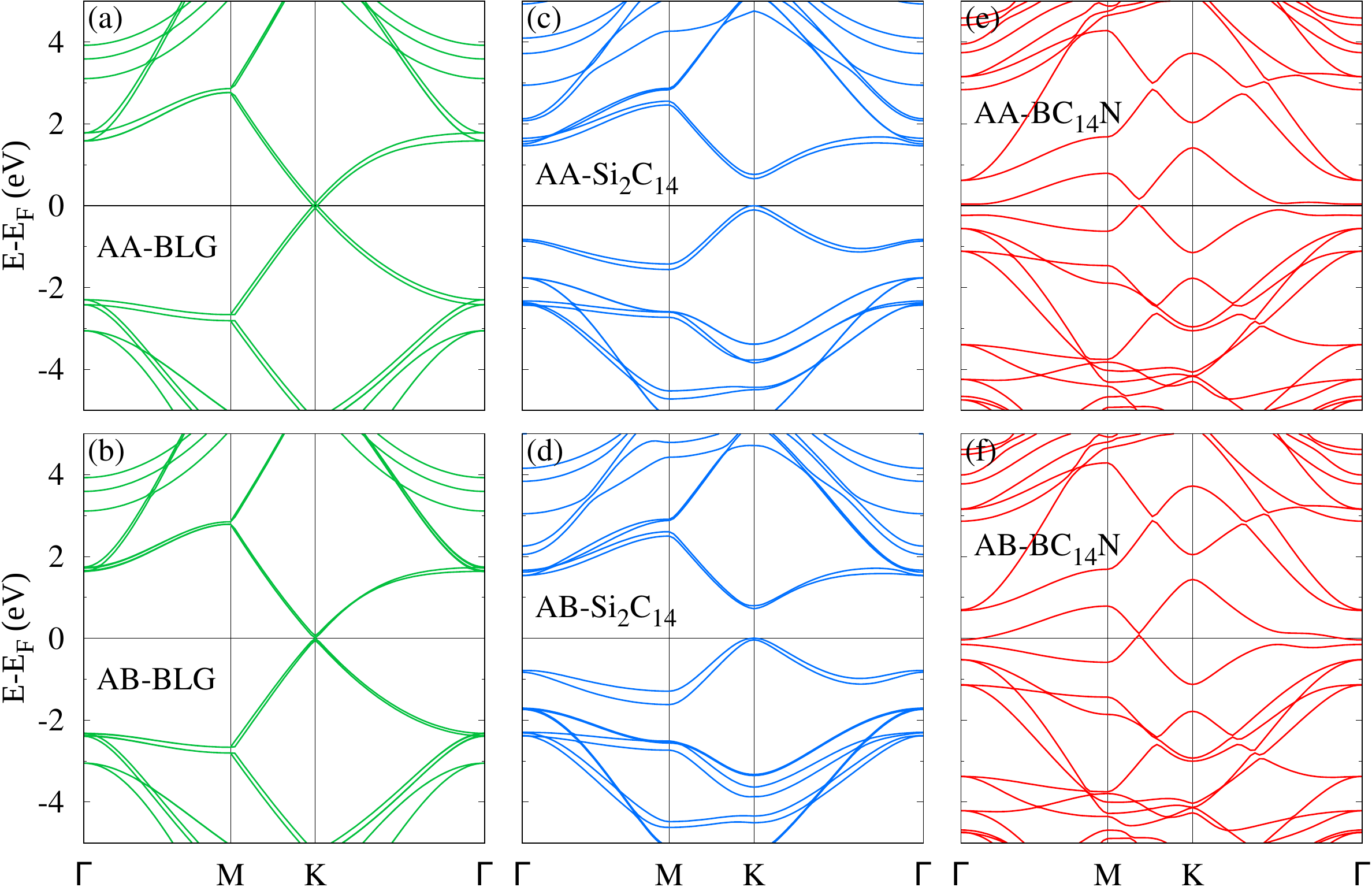}
	\caption{Electronic band structure of AA-stacked BLG (a), Si$_2$C$_{14}$ (b) and  BC$_14$N (c).
		Fermi energy is set at 0 eV }
	\label{fig02}
\end{figure}

The electronic band structure is plotted in \fig{fig02} for an AA-stacked BLG (a), Si$_2$C$_{14}$ (c), and BSi$_{14}$N (e), and AB-stacked BLG (b), Si$_2$C$_{14}$ (d), and BSi$_{14}$N (f).
In the AA-stacked BLG, the electron distribution pattern has a hexagonal symmetry which is almost the same as for a monolayer graphene. It thus generates a linear dispersion
of a valence band ($\pi$) and conduction band ($\pi^*$) intersecting the at K-point.
In the AB-stacked BLG, parabolic bands are found near the
Fermi level due to the asymmetric interactions between the upper
and the lower layers. The lowest valence band and the conduction band only have a weak overlap near the K-point. 
It means the density of free carriers is very low.
In both the AA- and the AB-stackings, the $\Gamma$-point determining the $\pi$-band width corresponds to the maximum 
and minumum $\pi$-band energy levels. The energy states are affected by 
interlayer atomic interactions at the M-ponit \cite{PhysRevB.74.085406}.

Furthermore, the energy spacing the between $\pi_2^*$ and the $\pi_1^*$ near the K-point is $0.11$, and $0.07$~eV for the AA- and the AB-stackings, respectively, which are equal to the corresponding energy spacing of the $\pi_2\text{-}\pi_1$, where $\pi_2$ and $\pi_1$ refer to the double states of the $\pi$-bands.
The linear dispersion of the AA-stacked BLG has been experimentally 
confirmed \cite{Kim2013}. 

The dopant atoms in BLG induce a symmetry breaking and the interlayer interactions result in a bandgap. 
Band gaps also appear due to the interactions of dopants arising from their lateral periodicity.
We therefore see a bandgap at the the K-point for the AA- and AB-stacked Si$_2$C$_{14}$ and AA-stacked BSi$_{14}$N \cite{RASHID2019102625}. 
The repulsive interaction in the AA- and the AB-stacked Si$_2$C$_{14}$ induces a direct bandgap which is $0.66$, and $0.72$~eV, respectively. Thus, both systems exhibit a semiconductor behavior. 
The attractive interaction in the AA-stacked BSi$_{14}$N generates a very small indirect bandgap, $0.025$~eV 
indicating semiconductor property. 
The B/N atoms change the position of the Dirac cone along the $M-K$ direction.
The Fermi energy crosses the valence band maxima and the conduction band minima in the AB-stacked BSi$_{14}$N 
revealing a degenerate semiconductor behavior. 
Furthermore, The energy spacing between the $\pi_2$ and the $\pi_1$
bands is increased in Si$_2$C$_{14}$, while the double states of the $\pi$-bands in 
BSi$_{14}$N totally disappears. This is in a good agreement with recent results for BN-codoped BLG in which 
the BLG is doped with one N atom in one layer and one B atom in the other \cite{Alattas2018}.

\subsection{Stress-Strain curves}
 
The mechanical properties of a structure can be determined by the stress-strain curve. A uniaxial tensile strain is applied to the atoms of both layers in the zigzag and the armchair directions. 
In the DFT calculation, the system is extended by small displacement increments, $0.02$, to the atoms at both ends. After each elongation, the system is relaxed to reach a new equilibrium
state with both ends fixed. The elongation and relaxation procedures are repeated until the desired tensile strain is reached. The same mechanism has been applied for BLG using MD simulation \cite{ZHANG20114511}.

\begin{table}[h]
	\centering
	\begin{center}
		\caption{\label{table_2}  The Young modulus (YM), The tensile strength (TS)m, 
		and the fracture stress (FS) of all the structures.}
		\begin{tabular}{|l|l|l|l|l|l|l|l|l|}\hline
			\multirow{2}{1.0cm}{AA}& \multicolumn{6}{p{0.1cm}}{\centering Zigzag}{\centering Armchair} \\
			\cline{2-7} & \multicolumn{1}{c|}{YM} & \multicolumn{1}{c|}{TS} & \multicolumn{1}{c|}{FS} & \multicolumn{1}{c|}{YM} & \multicolumn{1}{c|}{TS} & \multicolumn{1}{c|}{FS}          \\ \hline
			BLG	            & 974   &  99.64  &   99.64  &  974   &  96.22 & 96.22   \\
			Si$_2$C$_{14}$  & 732   &  50.22  &   46.2   &  728   &  36.72 & 36.72   \\		
			BC$_{14}$N      & 946   & 80.06   &   66.63  &  948   &  76.13 &  76.13  \\ \hline
			\multicolumn{1}{| c }{AB}                                                \\ \hline
			BLG	            & 885   &  90.58 &  90.58    & 882   &  88.5  &  88.5    \\
			Si$_2$C$_{14}$  & 665   &  45.63 &  38.1     & 670   &  42.55 &  39.34   \\		
			BC$_{14}$N	    & 898   &  72.71 &  41.9     & 855   &  71.72 & 71.72    \\ \hline
		\end{tabular}
	\end{center}
\end{table}

\begin{figure}[htb]
	\centering
	\includegraphics[width=0.45\textwidth]{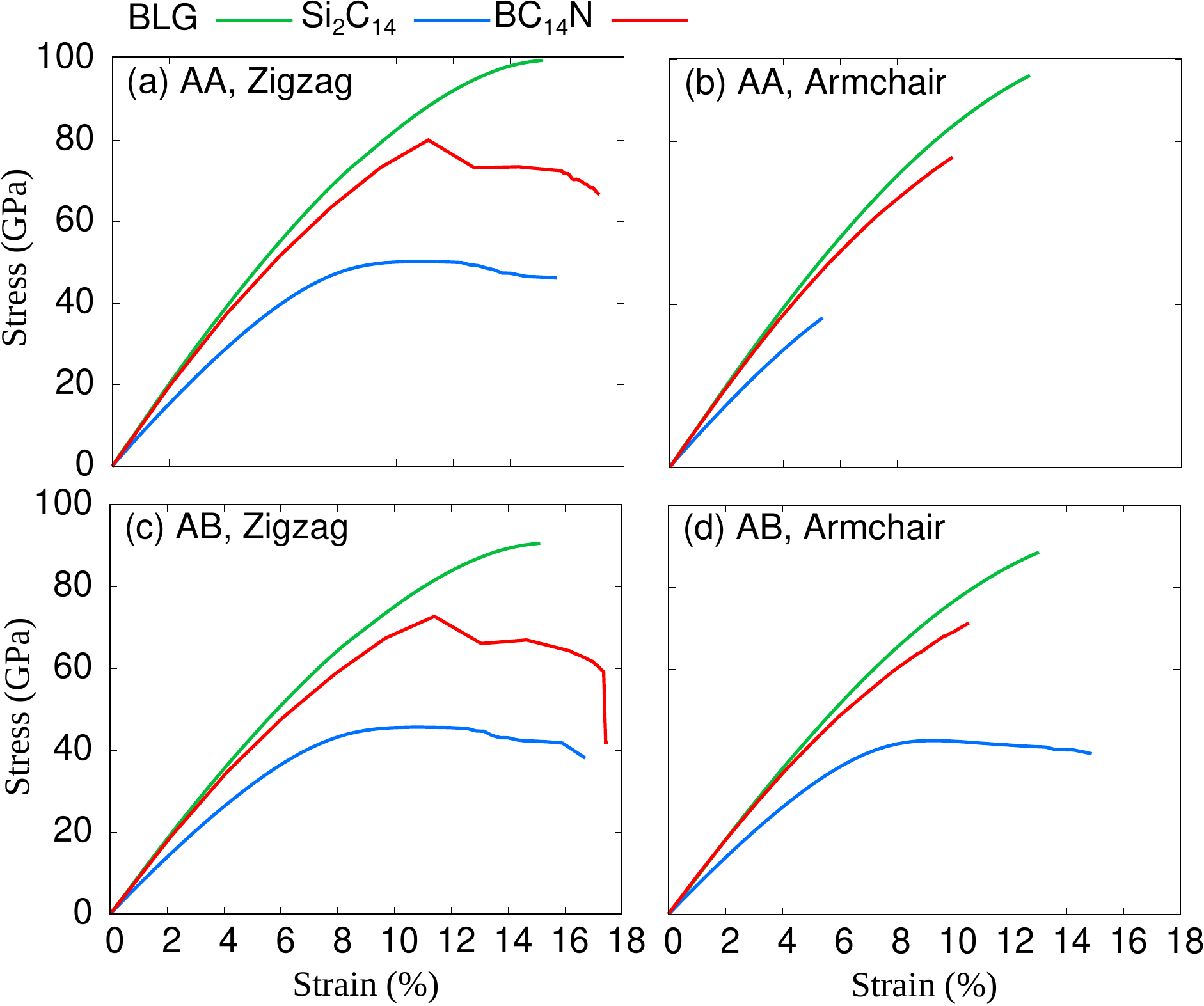}
	\caption{Stress-strain curves of AA- and AB-stacked BLG (green), Si$_2$C$_{14}$ (red), and  BC$_{14}$N (blue) in the zigzag (a, c) and armchair (b, d) directions.}
	\label{fig03}
\end{figure}

The stress-strain curves are shown in \fig{fig03} for both the AA- and the AB-stacked BLG (green), Si$_2$C$_{14}$ (blue), and  BSi$_{14}$N (red) in the zigzag (a and c) and the armchair (b and d) directions. 
In a pure AA-stacked BLG, the tensile strain linearly increases in a small regime of strain in which 
the system experiences enlargments linearly up to $\le 5\%$.
This linearity is measured by the Young moduli, which is
calculated as the initial slope of the stress–strain curve. The linear part of the curves is also called the elastic region. Table \ref{table_2} presents the Young modulus (YM), the tensile strain (TS), and the fracture strain (FS) of undoped and doped BLG. The first column in \tab{table_2} shows the structures, 
the column 2 to 4 are the values in the zigzag direction, and the last three column are the values 
in the armchair direction.  
The  stress–strain relationship of the AA- and the AB-BLG reveal a Young modulus of $974$ and $864$ GPa, 
respectively, which are very close to the Young moduli obtained in reports of experimental \cite{Lee385} and theoretical \cite{doi:10.1063/1.4789594} work.
At a large strain, the stress of the system respondes non-linearly to the strain until the system’s failure, 
determining the fracture strain.
The ultimate tensile strength is defined as the maxima in the stress-strain curve. 
The corresponding strain is introduced as the ultimate tensile strain indicating the flexibility of the system. 
The ultimate stress or tensile strain of the AA- and the AB-BLG is $99.64$ and $90.58$~GPa, respectively, 
at the strain of $0.151$. 
The values of ultimate stress are also equal to the fracture strain here as there is no stretching 
in the structure after the fracture strain~\cite{doi:10.1063/1.5091753}. 
The same behavior of the stress-strain curve is found for the armchair direction, 
where the Young modulus is the same for both the AA- and the AB-stackings.

In the AA- and the AB-stacked Si$_2$C$_{14}$ and BC$_{14}$N the stress-strain curves are modified due to 
the presence of dopant atoms. The bond energies of C-Si, C-B and C-N are smaller than those of C-C \cite{C2NR11728B, JAVVAJI201725}. The bonds can be generically arranged from high to low energy as: C-C, C-N, C-N and Si-C. We therefore see a reduction in the stress-strain curves of Si$_2$C$_{14}$ and BC$_{14}$N which are caused by the small bonding energies of the dopant atoms with the C atoms~\cite{C2NR11728B}. In addition, the super-cell expansion due to the dopant atoms mentioned before is another reason why the Si$_2$C$_{14}$ and BC$_{14}$N bilayers need less tensile stress.

Even though the repulsive interlayer interaction is increased in both the AA- and AB-stacked  Si$_2$C$_{14}$ the Young modulus still decreases. This indicates that the binding energy between C and Si atoms is dominant in controlling the elastic properties of the system. The same scenario can be applied for BC$_{14}$N where an attractive interaction between the layers exists.

Another observation is that the doped systems experience enlargments in a linear fashion till $\le 4\%$ 
in the AA-stacking, which demonstrates less elastic properties compared to the AA-stacking in both the zigzag 
and the armchair directions (see \tab{table_2}).

\subsection{Optical absorption spectra}

\begin{figure}[htb]
	\centering
	\includegraphics[width=0.48\textwidth]{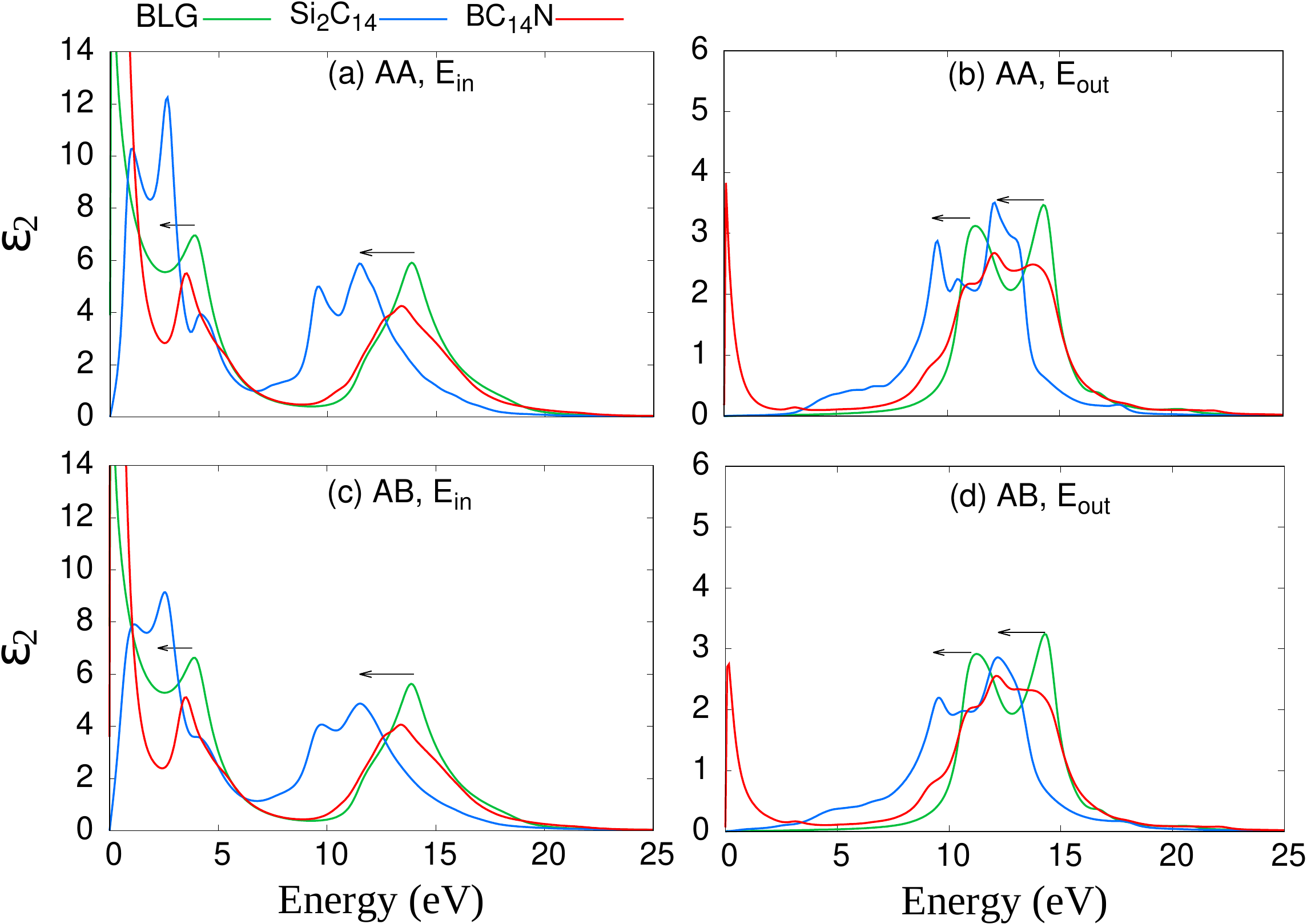}
	\caption{Imaginary part of dielectric function for the AA- (a and b) and the AB-stackings (c and d) in the case of an in-plane, E$_{\rm in}$, (left panel) and an out-of-plane, E$_{\rm out}$, (right panel) electric field}
	\label{fig04}
\end{figure}

In general, bilayer graphene has much richer spectral features in comparison with monolayers. 
We therefore present here the optical response of the AA- and the AB-stacked systems.
We find that the optical response of the dielectric function of a BLG can
be manipulated by the influence of repulsive or attractive interlayer interactions.
The imaginary part of dielectric function $\varepsilon_2$ is shown in \fig{fig04} for the AA- (a and b) and 
the AB-stackings (c and d) in the case of 
an in-plane, E$_{\rm in}$, (left panel) and an out-of-plane, E$_{\rm out}$, (right panel) electric field.

It well known that in the case of E$_{\rm in}$ the AA-stacked BLG has two main peaks in dielectric function at $3.95$ and $13.87$~eV formed by the $\pi$ to $\pi^*$ and the $\sigma$ to $\sigma^*$ transitions, respectively. 
Furthermore, in E$_{\rm out}$ the two main peak are generated by transitions from the $\sigma$ to $\pi^*$
at $11.22$~eV and the $\pi$ to $\sigma^*$ at $14.26$~eV.
The anisotropic behaviour is clearly observed for the two different polarizations \cite{NATH2015691}.
The AB-stacked BLG has the peaks at almost the same energy values with less intensity since it's interlayer 
distance is close to the AA-stacking. 

In the AA- and the AB-stacked Si$_2$C$_{14}$ two main features in $\varepsilon_2$ are observed in the case of 
E$_{\rm in}$. First, double peaks for each $\pi \rightarrow \pi^*$ and $\sigma \rightarrow \sigma^*$ transitions appear. This is attributed to increased energy spacing between the $\pi_{1,2}$, the $\pi^*_{1,2}$, the $\sigma_{1,2}$, and the 
$\sigma^*_{1,2}$ as is shown in \fig{fig02}(c and d). Second, a red shift towards lower energy for both peaks is seen. The red shift of the peaks occurs by decreased energy spacing between the $\pi$ and the $\pi^*$, and the $\sigma$ and the $\sigma^*$ a long the $\Gamma-M$ and the $M-K$ directions. It is interesting to note that the peak intensity at lower energy is enhanced. The peak intensity for the AA-stacking is higher than that of the AB-stacking. 
The peaks of Si$_2$C$_{14}$ in the case of out-of-plane electric field are not red shifted and 
the intensity of the peaks is almost the same for the AA-stacking, while it is slightly decreased for 
the AB-stacking.

The properties of the imaginary dielectric function for BC$_{14}$N is very different for both the in- and 
the out-of-plain electric fields. The intensity of the peaks is decreased due the overlaping of the valence and the conduction bands.
In the in-plane electric field, double peaks for both the AA- and the AB-stackings are not seen any more because 
of the absence the double states in the band structure. It also seems that the peak intensity is 
almost the same for the AA- and the AB-stackings. 
In the case of the out-of-plane electric field, the right peak is red shifted for the AA- and the AB-stackings 
while the left peak is diminished. In addition, a strong peak at a very low energy is observed. 
These are attributed to extreme decreasing of the energy spacing between the $\pi$ and the $\pi^*$, and the $\sigma$, and the $\sigma^*$ along the $\Gamma-M$ and the $M-K$ directions (see \fig{fig02}(e and f)).

\subsection{Seebeck coefficient}

\begin{figure}[htb]
	\centering
	\includegraphics[width=0.45\textwidth]{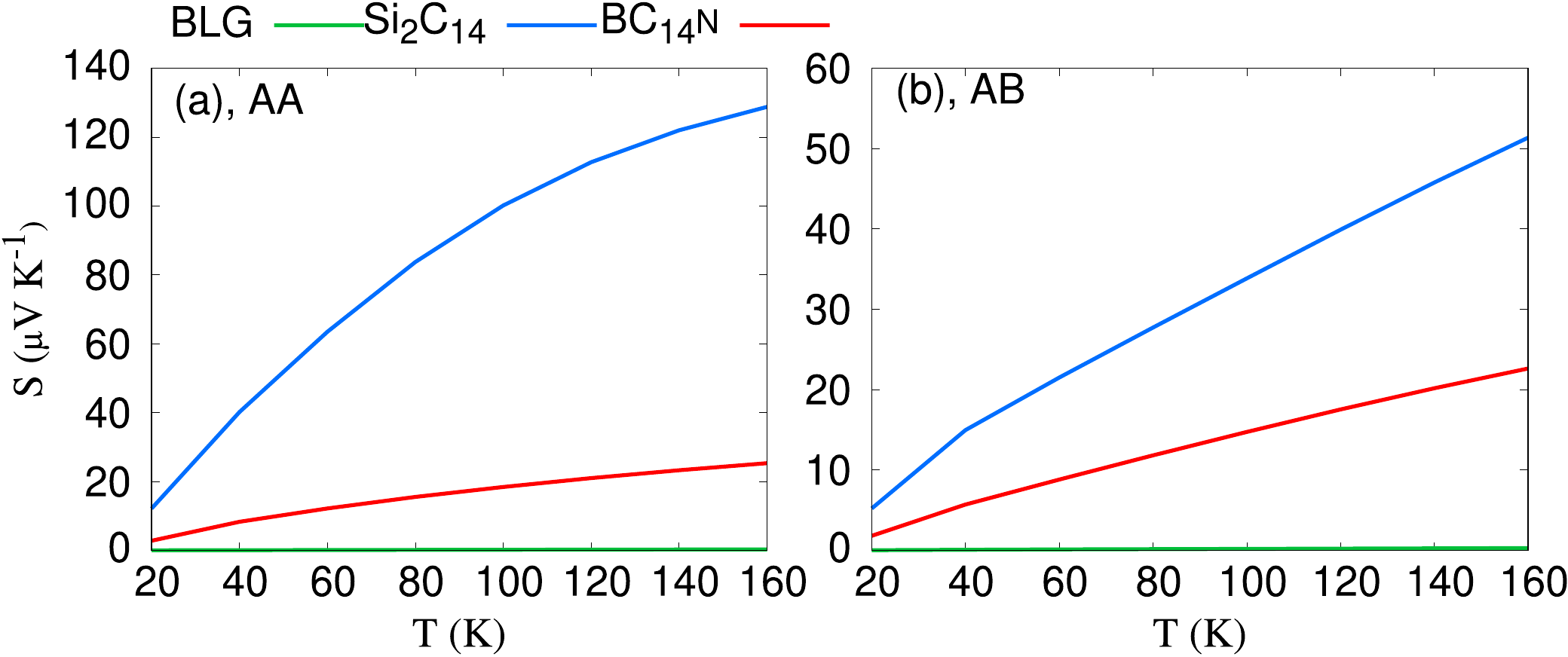}
	\caption{Seebeck coefficient for the AA- (a) and the AB-stacking (b), for pure BLG (green), Si$_2$C$_{14}$ (blue), and BC$_{14}$N (red).}
	\label{fig05}
\end{figure}

We investigated the thermal properties of our model at the low temperature ranging from $20$ to $160$~K, where the phonons are not active \cite{PhysRevB.87.241411,ABDULLAH20181432}. So, the electrons deliver the main contribution to the thermal behavior. It is know that a good thermoelectric material
should have a high electrical conductivity, Seebeck coefficient, $S$, and low thermal conductivity.
The thermoelectric performance of monolayer and bilayer graphene is poor because of closed bandgaps, 
leading to a small Seebeck coefficient \cite{DENG2019622, Abdullah_2018}. This can be clearly seen in \fig{fig05}, 
where the $S$ versus temperature is plotted for both the AA- (a), the AB-stacking (b), for pure BLG (green), Si$_2$C$_{14}$ (blue), and BC$_{14}$N (red). The Seebeck coefficient is very small for pure BLG.

The key point to enhance the $S$ is the opening up of the bandgap. Since the bandgap in Si$_2$C$_{14}$ is much 
larger than that of BC$_{14}$N, a much higher $S$ of Si$_2$C$_{14}$ is observed for both  the AA- and the AB-stackings. Therefore, one can expect to have a higher thermoelectric performance for the Si$_2$C$_{14}$ structure. 
The electronic and thermal properties of doped BLG may be of fundamental
interest and can play an important role in the performance of nanoscale devices.\\

\section{Conclusion}~\label{Sec:Conclusion}
We investigate the interaction energy, the electronic band structure, the mechanical, the optical and 
the thermal characteristics of AA- and AB-stacked BLG, Si$_2$C$_{14}$, and BC$_{14}$N structures. 
We show that interlayer interaction effects have a crucial influence on the properties of both the Si$_2$C$_{14}$, and the BC$_{14}$N structures. The Si atoms in Si$_2$C$_{14}$ and the B and N atoms in BC$_{14}$N induce repsulsive and attractive interactions between the layers, respectively. We find that the influence of the repulsive interactions 
on the stress-strain curves in Si$_2$C$_{14}$ is larger compared to the attractive interactions in BC$_{14}$N.
Therefore, the stress-strain curves are more suppressed for 
Si$_2$C$_{14}$ in both the zigzag and the armchair directions. The imaginary dielectric function of Si$_2$C$_{14}$ induces a red shift at high energy while only a reduction in the imaginary dielectric function is found for BC$_{14}$N without any important shift of peaks. 
Furthermore, we find that Si$_2$C$_{14}$ has a more promissing thermal response than BC$_{14}$N with regards to
possible applications in devices.

\section{Acknowledgment}
This work was financially supported by the University of Sulaimani and 
the Research center of Komar University of Science and Technology. 
The computations were performed on resources provided by the Division of Computational 
Nanoscience at the University of Sulaimani.  
 
%\section{References}

%\bibliographystyle{elsarticle-num} 
%\bibliography{Ref_2.bib}

\end{document}